\documentclass[amsmath,amssymb,aps,prb,twocolumn,groupedaddress,floatfix]{revtex4-1}

\usepackage{natbib}
\usepackage{revsymb4-1}
\usepackage{graphicx}

\begin{document}


\title{Topographic phase boundary shifts and saturation for
  anisotropic ion straggle during sputter etching}


\author{Emmanuel O. Yewande}
\email[Corresponding author: ]{eo.yewande@ui.edu.ng}
\author{Raphael O. Akande}
\affiliation{Theoretical Physics Group, Department of Physics, Faculty of Science, University of Ibadan, Ibadan, Nigeria.}


\date{\today}

\begin{abstract}
Surfaces sputtered by ion beam bombardment have been known to exhibit
patterns whose 
behavior is modeled with stochastic partial differential equations. A
widely 
accepted model is the Cuerno-Barabasi model which is robust in its
predictions 
of sputtered surface morphologies. An understanding of the factors
responsible for such 
surface topographies can be achieved by using scaling arguments on the
stochastic 
model. For such explanations, knowledge of the coefficients is
crucial. The 
more so since these vary with different materials, the sputtering
process itself 
generates non-equilibrium surfaces within some finite timescale, and
the implication of 
recent results of surface topographies unexplained by the continuum theory. We calculate and study these coefficients as functions of the sputtering parameters for yet unreported cases of anisotropic ion energy distribution within the sputtered material. Consequently, we present phase diagrams for the significant case of anisotropic ion straggle. We observe shifts in the phase boundaries when the collision cascade geometry rotates, and we also found saturation behavior in the diagrams; in which case the boundaries become independent of the penetration depths. Our results indicate a possible origin of yet unexplained nanodot topographies arising from oblique incidence ion etching of amorphized surfaces.

\end{abstract}

\pacs{}

\maketitle

\section{Introduction}
Surface sputtering is a process by which materials are removed from
the surface of a solid through the impact of energetic
particles. There are many sputtering techniques, some of which are for
industrial uses. It is a widely applicable technology with remarkable
level of sophistication. Sputtering can be used for surface analysis,
depth profiling, surface cleaning, micromachining, deposition, surface
coating, semiconductor doping, etching, magnetic storage technology,
design of nanostructures on a surface and many more.\cite{Carter1983,
  Yewande_PhDthesis, Munoz-Garcia2007, Makeev2002} 

The sputtering process creates patterns on the surface at nanometer
length scales; hence 
there is the need to study the surface morphology for knowledge and
control of the pattern 
of these nanostructures exhibited on the surface when it is
sputtered. The periodic 
ripple patterns observed for off-normal incidence sputtering were
found to have either 
of two orientations, depending on the angle of incidence of the
impinging energetic ions 
bombarding the surface. \cite{Eklund1993, Carter1983, Eklund1991,
  Habenicht1999, Habenicht2002, Krim1993} The ripples are oriented
perpendicular to 
the projection of the ion-beam direction onto the surface plane for
small 
incidence angles, whereas they are oriented parallel to this
projection for 
grazing angles. However, a number of experiments on surface sputtering
have observed 
only a kinetic roughening of the surface, or nanodots, instead of the
periodic 
ripple patterns observed by others. \cite{Chason1994, Mayer1994,
  Gago2001, Facsko2001, Frost2000, Ziberi2006} 

Bradley and Harper (BH) were the first to theoretically account for
the 
ripple morphology on amorphous substrate using the linear continuum
theory, \cite{Bradley1988} which entails the theoretical modeling of
the 
surface as a continuum of points whose evolution with time is governed
by a 
deterministic partial differential equation. Their theory provides an
understanding of ripple formation and dependence of orientation on ion
incidence, but predicts an exponential
increase of the ripple amplitude. Experiments, however, observed rough
surfaces at larger lengthscales, and a
saturation of the surface width. \cite{Chason1994, Mayer1994} 

An improvement of the linear BH theory is the widely acceptable
Cuerno-Barabasi (CB) 
model which is robust in its predictions of sputtered surface
morphologies. \cite{Makeev2002, Cuerno1995} CB model is a stochastic
nonlinear 
continuum model which considers nonlinearities in addition to the
linear theory. 
CB derived this stochastic non-linear equation, that mimics the
surface evolution 
with respect to time, by working in the laboratory frame of reference,
such that 
the coefficients appearing in the equation are functions of the actual
experimental 
sputtering parameters. An understanding of the factors responsible for
such 
surface topographies predicted by CB can be achieved by using scaling
arguments 
on the stochastic model based on the relative signs of the
coefficients; which are the weighting factors in the continuum model. \cite{Cuerno1995} 

However, experiments by Fackso and co-workers on normal incidence
sputtering of 
GaSb \cite{Facsko1999} and Si \cite{Gago2001, Castro2005} observed nanodots
instead of ripple 
morphologies or kinetic roughening. Moreover, recent results of the
discrete theory in Ref. \onlinecite{Yewande2006} and \onlinecite{Yewande2007}, from Monte Carlo
simulations of 
the time evolution of the discretized surface, predicts the presence
of such nanodots 
for oblique incidence sputtering as well. This has been recently
confirmed by 
experiments. \cite{Ziberi2006} Much is still unclear about the
continuum theoretical explanations of these recent topographies. 
Thus, knowledge of the surface topographic 
weighting factors, of the
stochastic partial differential equation that describes the surface
evolution, and their comparative analysis is crucial for an understanding and
explanation of the 
predicted surface topographies and morphology and for knowledge of how
these differ from the continuum theory predictions. Since the
continuum theory 
calculations have, however, not yet been performed for values of the sputtering
parameters of the order of those
used in the simulations it is crucial that this be done in order to
ascertain whether, 
among other things, the nanodots predicted and observed for oblique 
incidence sputtering constitute a breakdown of the theory.  

Hence, in this paper we calculate and comparatively study these
sputter coefficients as functions of the sputtering parameters for the case of anisotropic
distribution of the energy of the impinging ion within the sputtered
material. In which case, we have performed calculations for values of
the order of those in the simulations. But, in this first instance, we
restrict ourselves mainly to a comparison of the relative strengths of
the anisotropic cascade parameters.Our results indicate shifts and
saturation of phase boundaries due to the relative strengths of the
perpendicular cascading indicators, and suggests a possible
origin 
of yet unexplained nanodot topographies arising from oblique 
incidence ion etching of amorphized surfaces.

The rest of this paper is organized as follows. In the next section we
discuss 
the linear and the nonlinear continuum theory. 
In section \ref{sec:results}, we present and discuss our results of the phase
boundaries for the 
yet unreported anisotropic cases. Finally, we give our conclusions in section \ref{sec:conclusions}.

\section{\label{sec:theory} Continuum Theory}
For clarity in what follows we provide a schematic diagram, of the ion
bombardment of the substrate in a sputtering process, below in Fig.\ \ref{fig:schematic}.

\begin{figure}
\includegraphics[angle=270, width=0.54\textwidth]{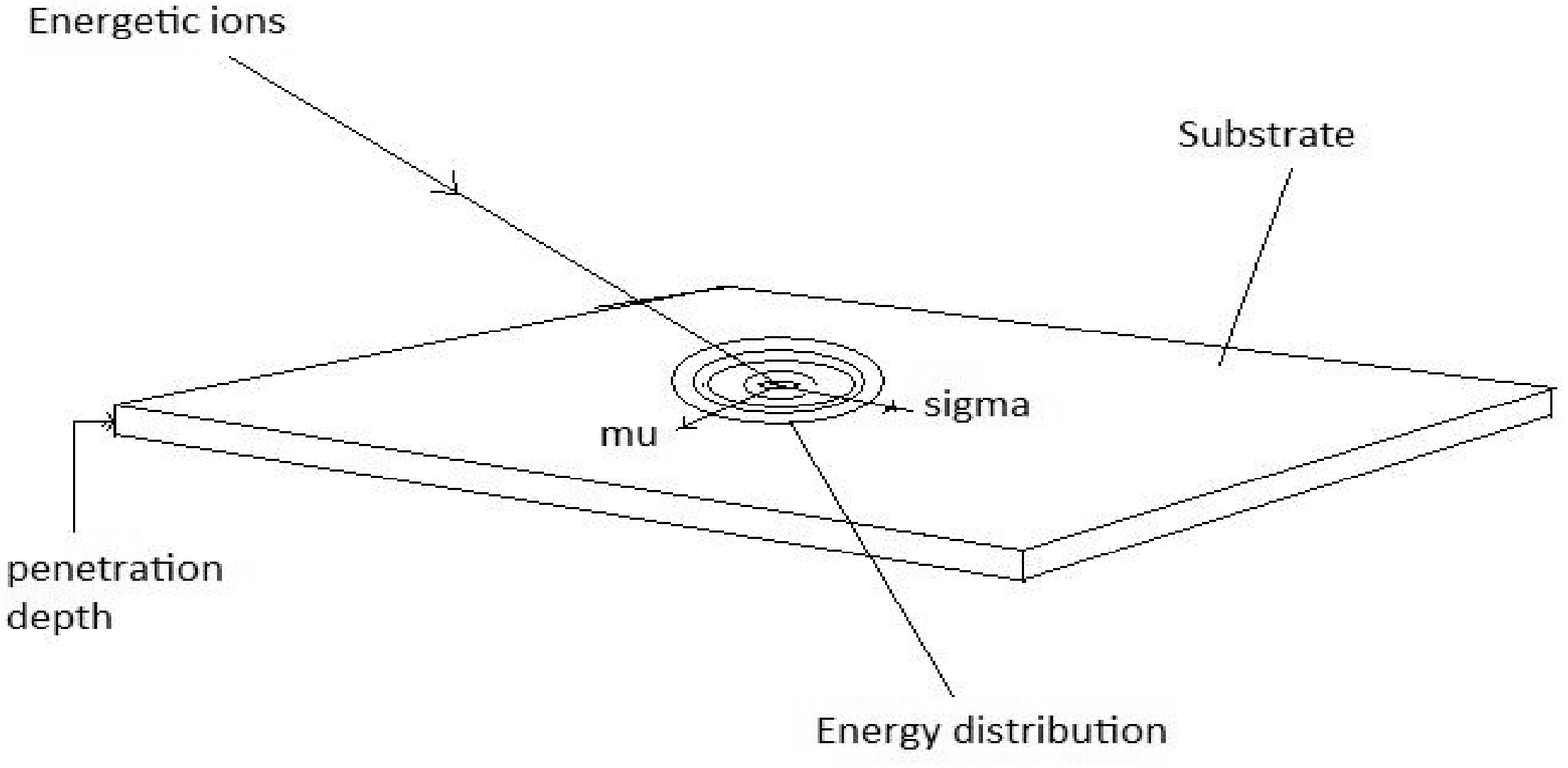}
\caption{\label{fig:schematic} A schematic diagram of the ion sputtering process.}
\end{figure}

The linear theory of BH \cite{Bradley1988} is given as Equation
\ref{eq:BH} while the non-linear theory of 
Cuerno Barab{\'a}si \cite{Cuerno1995} is given as
Eq. \ref{eq:CB}. Equation \ref{eq:CB} is an improvement on \ref{eq:BH}
with 
the introduction of the non-linear terms.

\begin{equation}
\label{eq:BH}
\frac{\partial h}{\partial t}=-\upsilon_0 + \upsilon_0\frac{\partial
  h}{\partial x} + \beta_x\frac{\partial^2 h}{\partial x^2} +
\beta_y\frac{\partial^2 h}{\partial y^2}
- D\nabla^2 \left(\nabla^2 h\right)
\end{equation}

\begin{eqnarray}
\label{eq:CB}
\frac{\partial h}{\partial t}=-\upsilon_0 + \gamma\frac{\partial
  h}{\partial x} + \nu_x\frac{\partial^2 h}{\partial x^2} +
\nu_y\frac{\partial^2 h}{\partial y^2} + \frac{\lambda_x}{2}\left(\frac{\partial
  h}{\partial x}\right)^2
\nonumber \\
 + \frac{\lambda_y}{2}\left(\frac{\partial
  h}{\partial y}\right)^2 - D\nabla^2 \left(\nabla^2 h\right) + \eta
\; \; \;
\end{eqnarray}

$h$ is the surface height at position ${\bf r}$ and time $t$, and
$\upsilon_0$ is 
the erosion velocity of a flat surface. $\beta_x$ (or $\nu_x$) is the
surface tension 
coefficient along the $x$ direction, which accounts for the
instability created 
by the sputtering process in which surface depressions (troughs) are
eroded 
in preference to protrusions (crests). $\lambda_x$ is the nonlinear
coefficient 
along the $x$ direction, which arises as an improvement of the linear
theory 
to account for the stability of the ripple amplitude and other
nonlinear effects 
observed in experiments. $D$ is the surface diffusion coefficient,
which 
takes the thermal hopping of surface atoms into consideration; such
migration 
of surface atoms is the counteracting process tending to restore the
surface 
tension instabilities. 

The surface patterns observed arises from the
right balance 
of the interplay between the surface tension instabilities arising
from the erosion 
process and the stabilizing effects of surface diffusion. $\eta$ is
the noise term 
representing the random nature of the sputtering of surface material;
it is 
taken to be a random variable with a Gaussian distribution and zero mean.       
The difference between $\beta_x$ and $\nu_x$ is that the former is a
function of the 
local sputtering parameters in the local surface frame of reference,
whereas the 
later is a function of the experimental sputtering parameters in the
laboratory 
frame of reference. Thus, apart from Equation \ref{eq:CB} being a more
realistic 
theoretical picture of the sputtering process, it is also the most
relevant in our 
investigations here for direct comparison with the experimental
conditions and, 
hence, results.
The coefficients occuring in  Equation \ref{eq:CB} (derived in
Refs.\ \onlinecite{Cuerno1995} and \onlinecite{Makeev2002}) are
provided 
for ease of reference in the rest of this paper as follows. 

\begin{widetext}
\begin{equation*}
\nu_x = \Upsilon
a\frac{a_\sigma^2}{2\varpi^3}\left(2a_\sigma^4\sin^4
  \theta-a_\sigma^4a_\mu^2\sin^2 \theta \cos^2 \theta
  +a_\sigma^2a_\mu^2\sin^2 \theta \cos^2 \theta -a_\mu^4\cos^4
  \theta \right),
\nu_y = -\Upsilon a\frac{\cos^2 \theta a_\sigma^2}{2\varpi}
\end{equation*}
\begin{eqnarray*}  
\lambda_x =
\Upsilon\frac{\cos \theta}{2\varpi^4}[a_\sigma^8a_\mu^2\sin^4 \theta
\left(3+2\cos^2 \theta \right)+4a_\sigma^6a_\mu^4\sin^2 \theta \cos^4
\theta
-a_\sigma^4a_\mu^6\cos^4 \theta \left(1+2\sin^2 \theta\right)] \\
-\varpi^2[2a_\sigma^4\sin^2 \theta
-a_\sigma^2a_\mu^2\left(1+2\sin^2 \theta \right) ] 
 -a_\sigma^8a_\mu^4\sin^2 \theta\cos^2 \theta-\varpi^4 
\end{eqnarray*}
\begin{equation}
\label{eq:coefficients}   
\lambda_y = \Upsilon\frac{\cos \theta}{2\varpi^2}\left(a_\sigma^4\sin^2 \theta +
  a_\sigma^2a_\mu^2\cos^2 \theta - a_\sigma^4a_\mu^2\cos^2 \theta -
  \varpi^2\right)
\end{equation}

\end{widetext}
where
\begin{eqnarray}
a_\sigma = \frac{a}{\sigma}, a_\mu = \frac{a}{\mu}, \varpi =
a_\sigma^2\sin^2 \theta + a_\mu^2\cos^2 \theta, \nonumber \\
\Upsilon =
\frac{FEJa}{\sigma\mu\sqrt{2\pi\varpi}}\exp\left(-a_\sigma^2a_\mu^2\cos^2
  \theta/2\varpi\right).
\nonumber
\end{eqnarray}
$F$ is the ion flux and $J$ is the proportionality constant
between the power deposition and the rate of erosion. For convenience
we have set $FEJ=1$, which implies the coefficients are in units of $FEJ$.
Cuerno and Barab{\'a}si presented results of phase diagram
calculations for the isotropic case for simplicity and convenience in
their introduction of the theory.\cite{Cuerno1995} They found three
different scaling regimes defined by the relative signs of the surface
tension and nonlinear coefficients. These scaling regimes are
applicable to an understanding of the large length scale (of the order
of tens of µm) topographies as found in experiments in which
topographies have been presented for 
lower resolution of the probe instruments [often atomic force
microscopes (AFM) or 
scanning tunneling microscopes (STM)]. 

In this paper we have
interpreted our 
results for the anisotropic case in terms of these three scaling
regimes when 
nonlinearities are relevant. Recent work in Ref.\ \onlinecite{Yewande2010}
found a 
much larger number of possibilities for the anisotropic case.
Following Cuerno and Barab{\'a}si, the regions to be encountered below
are defined by the signs of the linear and nonlinear coefficients as
follows - 
I: $\nu_x<0, \nu_y<0, \lambda_x<0, \lambda_y<0$; II: $\nu_x<0, \nu_y<0,
\lambda_x>0, \lambda_y<0$; III: $\nu_x>0, \nu_y<0,
\lambda_x>0, \lambda_y<0$. For regions I and II, orientations of the
ripples formed are 
along the $x$ direction as this presents the highest magnitude of the
modulus of 
the surface tension coefficients. Region III is characterised by
orientations along 
the $y$ direction since in this case $\nu_x>0$. 

\section{\label{sec:results} Results and Discussion}
The results presented here are obtained from numerical computations of
the equations 
of section \ref{sec:theory} for $a$  ranging between 0 and 3, and for
values of
 $\theta$ ranging between $0$ and $90^\circ$; where we obtain a set of
 values for each 
of $\nu_{x/y}$ and $\lambda_{x/y}$, and perform a comparative study of
their relative 
signs for the phase plots. We have not considered higher values of $a$
in order to be able to display the saturation behaviour, which we
found to depend on $a$, without considering values of the
collision cascade parameters higher than those in the simulations. 

\begin{figure}
\includegraphics[angle=270, width=0.45\textwidth]{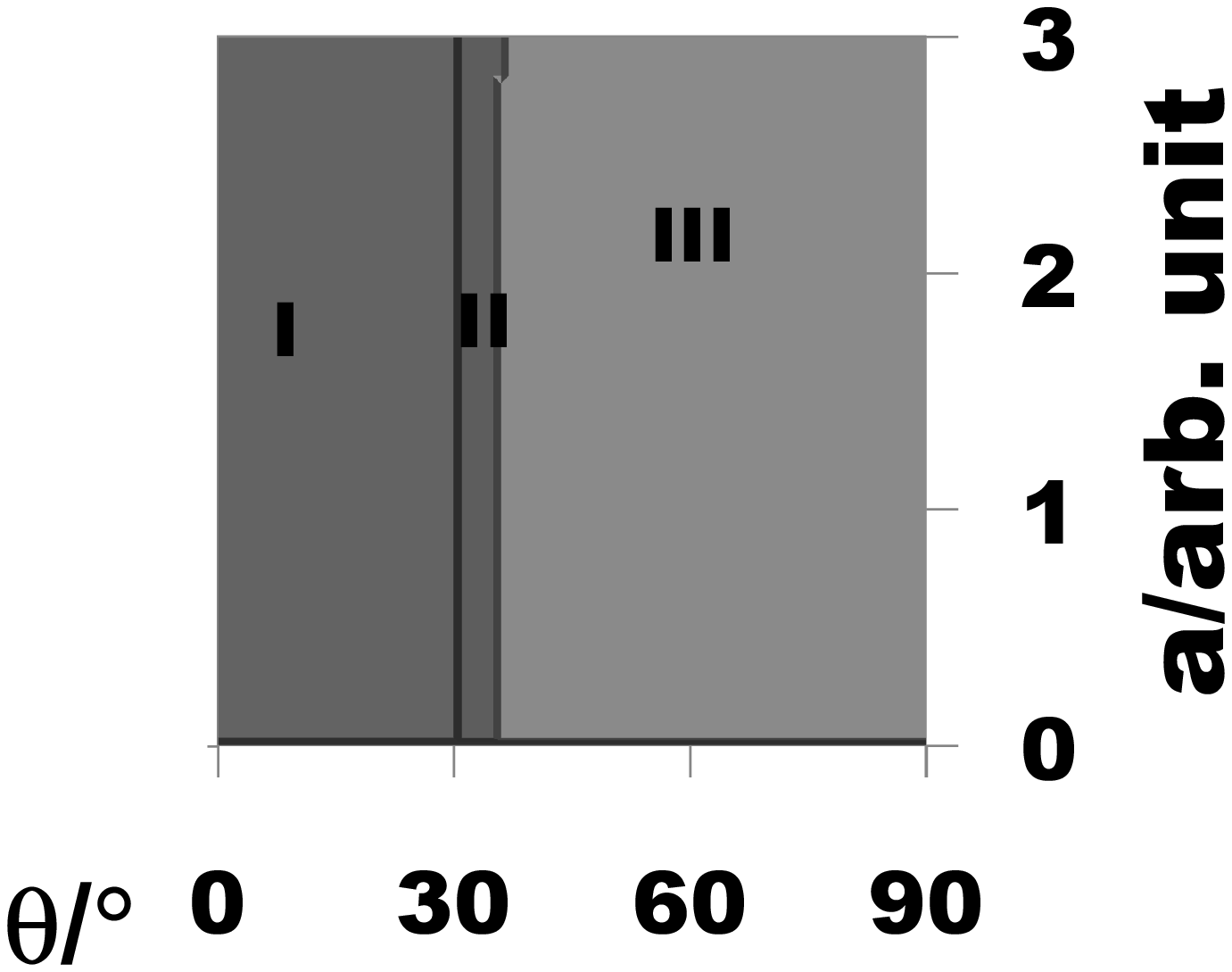}
\caption{\label{fig:isos7m7} Phase diagram for the isotropic case
  $\sigma = \mu = 7$. The phase boundaries are found to saturate. The
  regions are as 
defined and explained in the text.}
\end{figure}

Shown below are contour plots of the
obtained data with 
which we shall make comparison and then make conclusions. 
First, we consider higher values of $\sigma$ and $\mu$ that have not yet
been reported for 
the isotropic case introduced in Ref.\ \onlinecite{Cuerno1995}, and
found the 
saturation behavior presented in Fig.\ \ref{fig:isos7m7} below for
$\sigma = \mu = 7.0$. As can be seen from this figure, the phase
boundaries become 
independent of the penetration depth, $a$, for wider
ellipsoid of ion straggle within the substrate (see Fig.\ \ref{fig:schematic}).

\begin{figure}
\includegraphics[angle=270, width=0.45\textwidth]{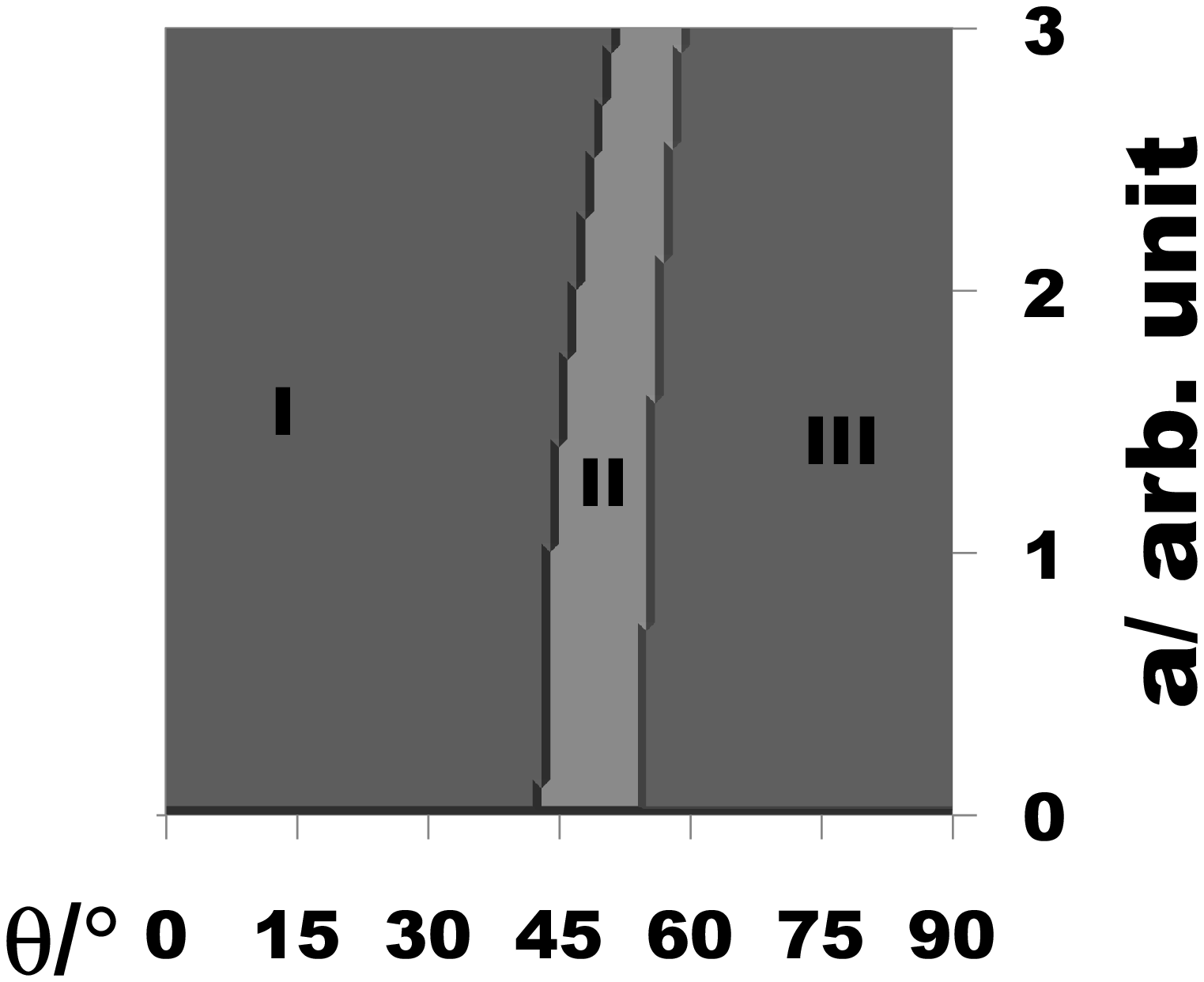}
\includegraphics[angle=270, width=0.45\textwidth]{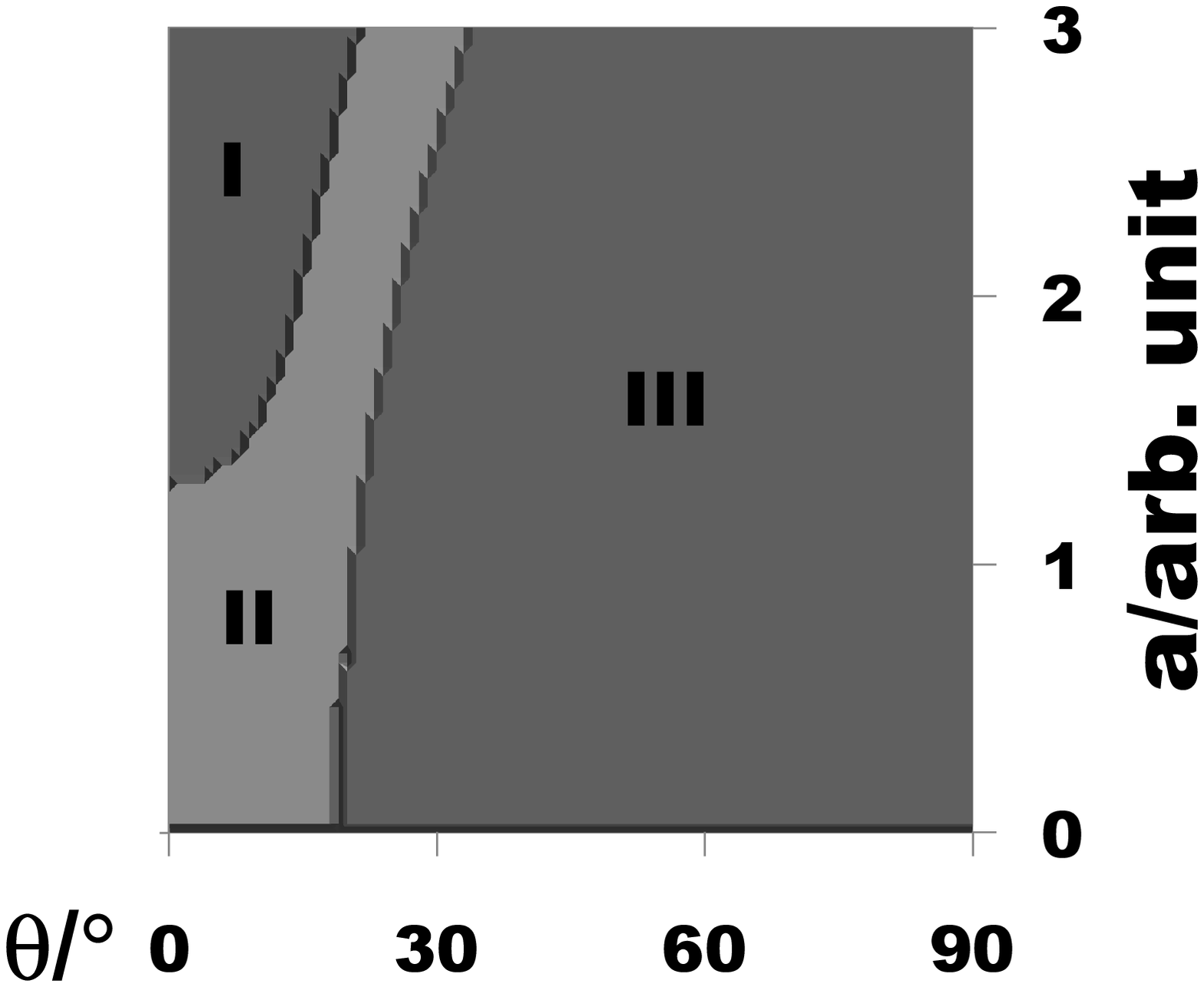}
\caption{\label{fig:lls_sgtm} Phase diagram for the anisotropic case
 $\sigma = 3.5$, $\mu = 1.5$ [top]; and vice-versa.}
\end{figure}

\begin{figure}
\includegraphics[angle=270, width=0.45\textwidth]{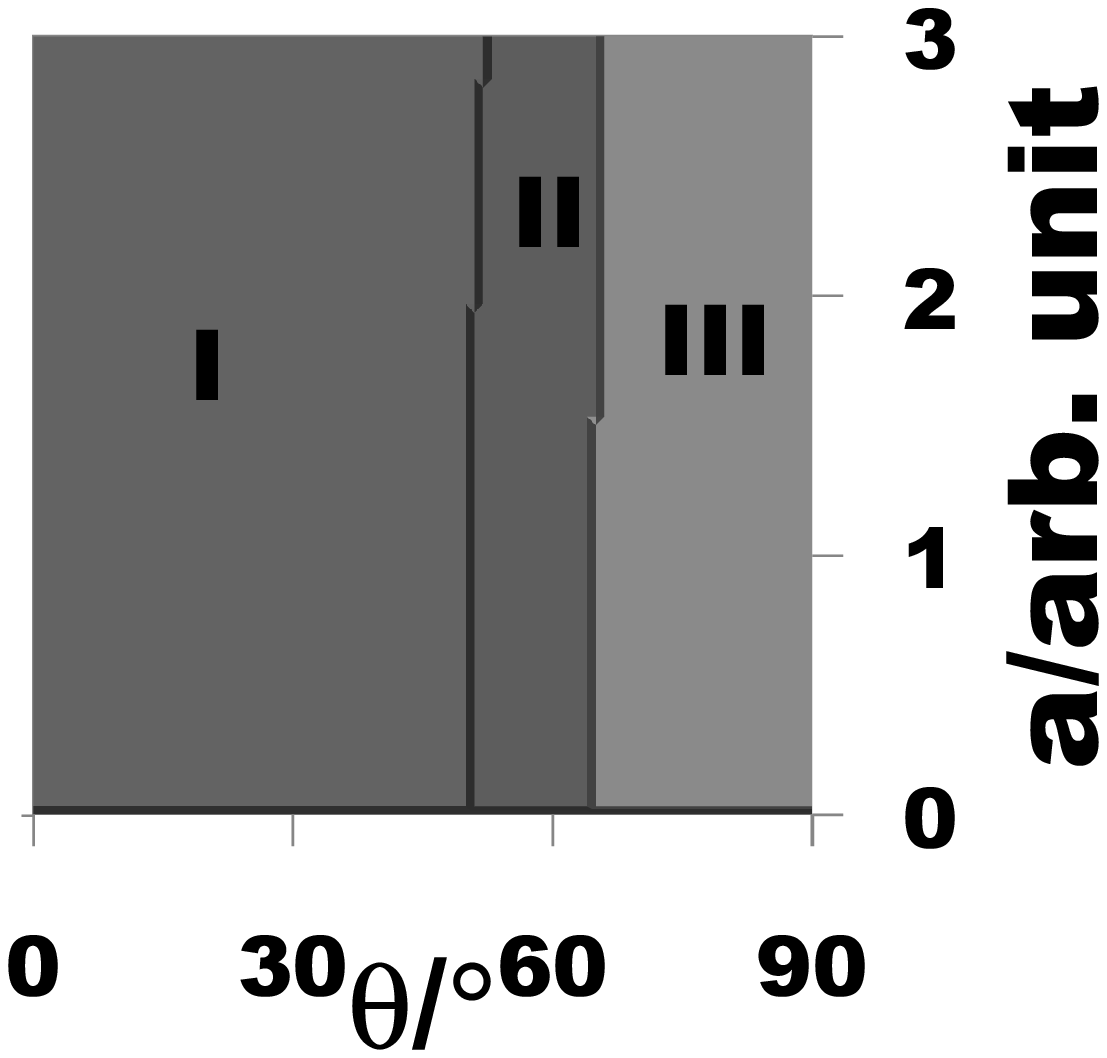}
\includegraphics[angle=270, width=0.45\textwidth]{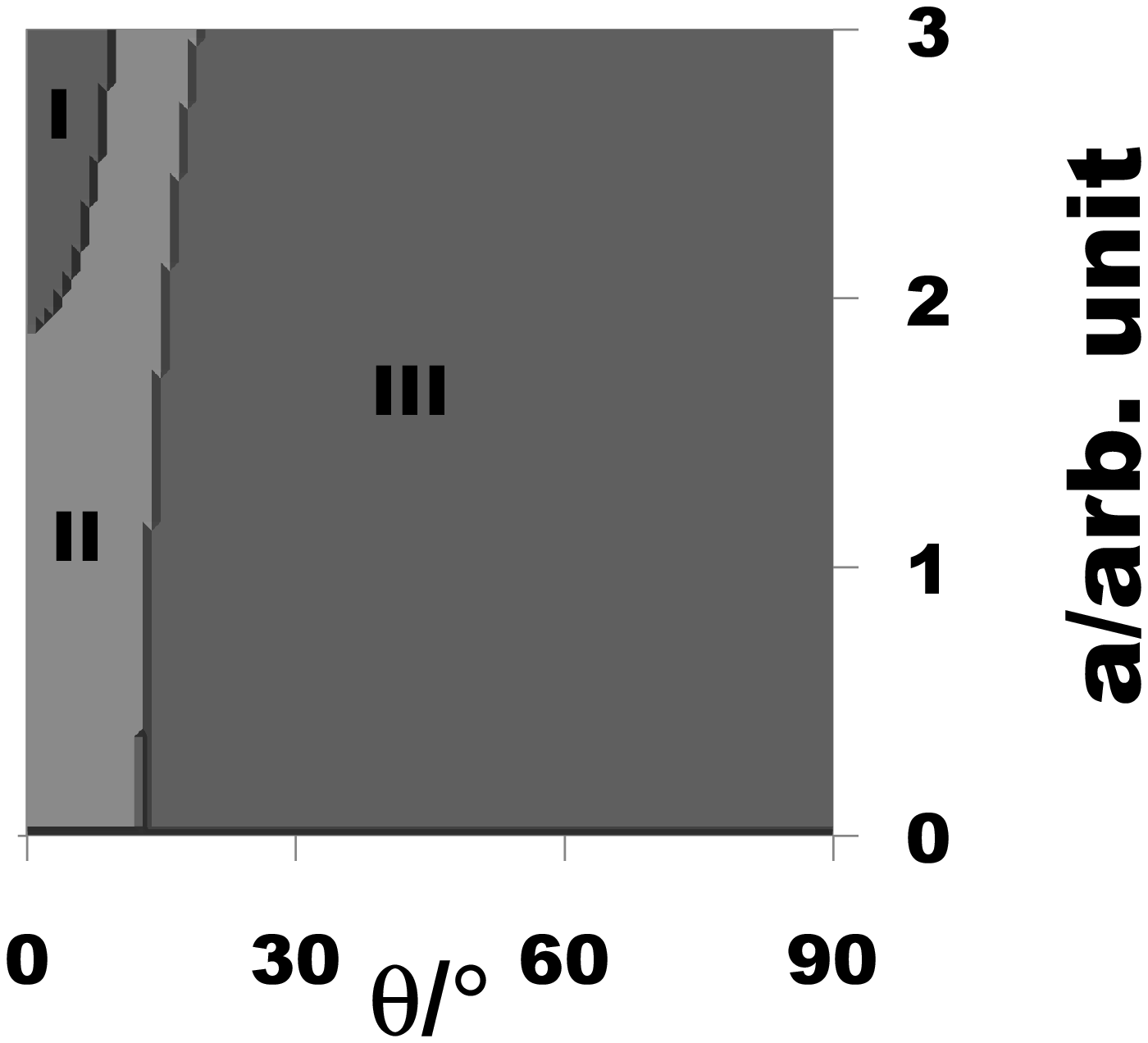}
\caption{\label{fig:lls_s6m2} Phase diagram for the anisotropic case
  $\sigma = 6$, $\mu = 2$ [top]; and vice-versa (b)}
\end{figure}

\begin{figure}
\includegraphics[angle=270, width=0.45\textwidth]{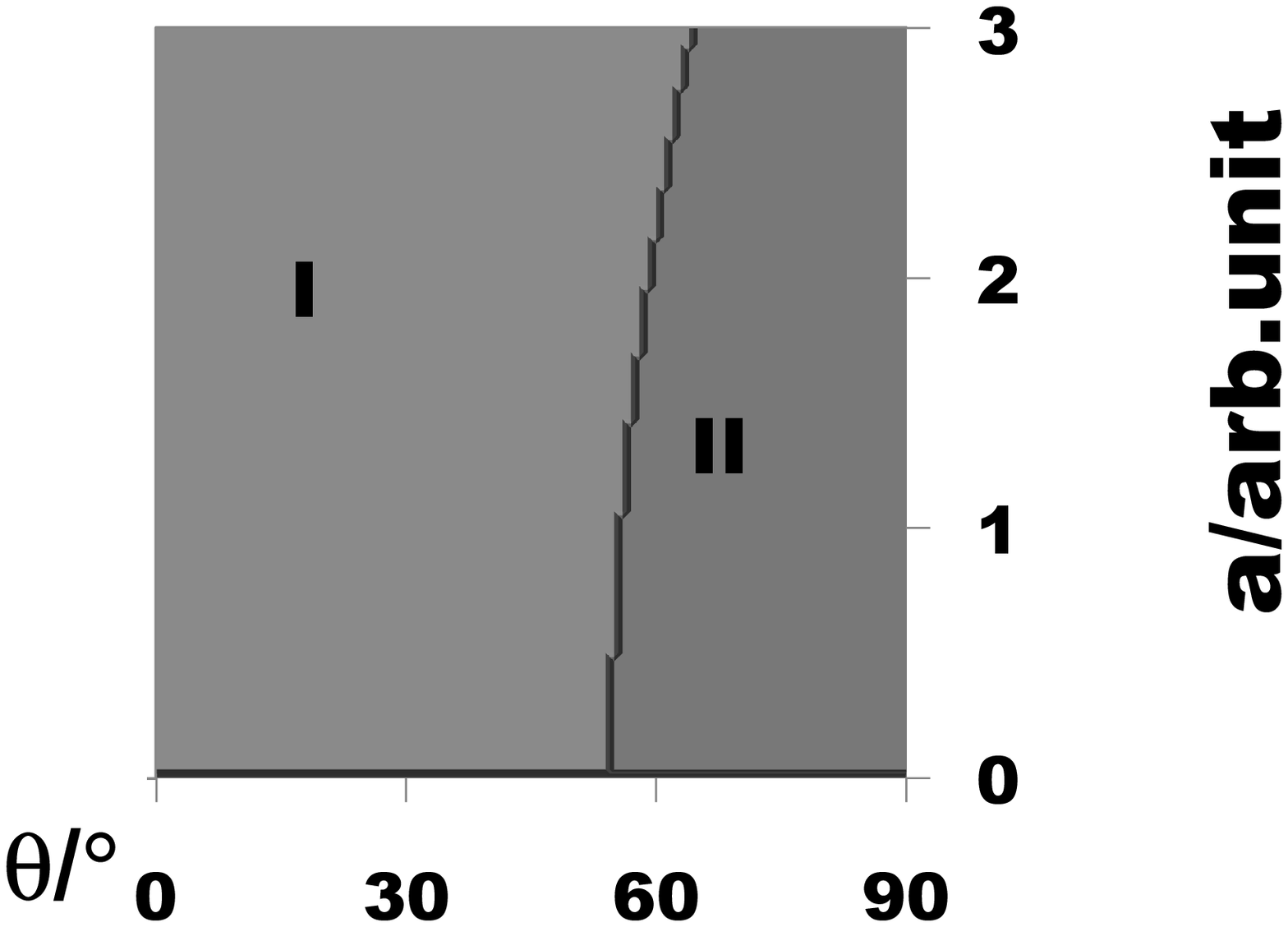}
\includegraphics[angle=270, width=0.45\textwidth]{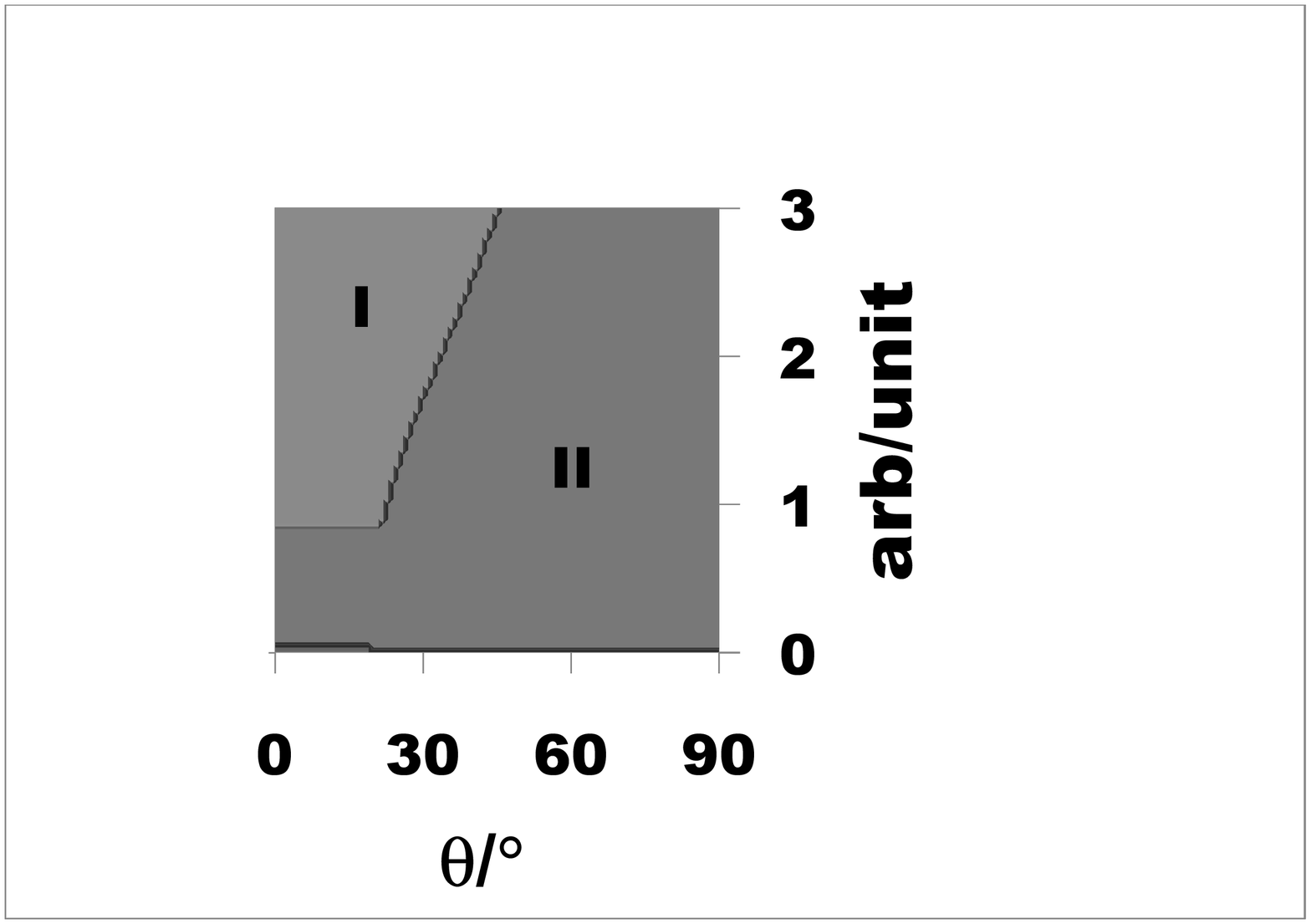}
\caption{\label{fig:sls_s1m2} Short length scale phase diagram for the anisotropic case
  $\sigma = 2$, $\mu = 1$ [top]; and vice-versa. The regions observed are as defined in the text.}
\end{figure}

\begin{figure}
\includegraphics[angle=270, width=0.45\textwidth]{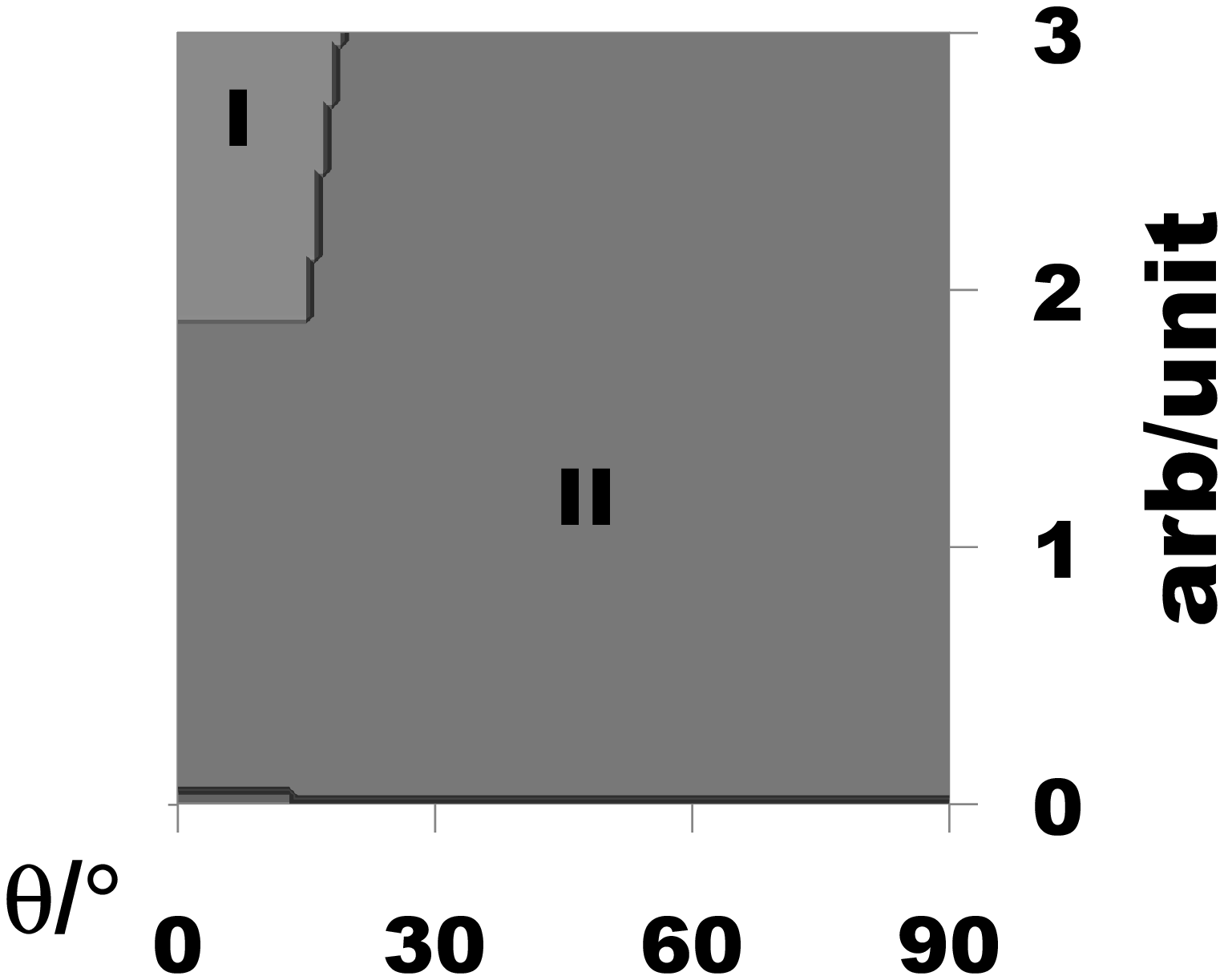}
\includegraphics[angle=270, width=0.45\textwidth]{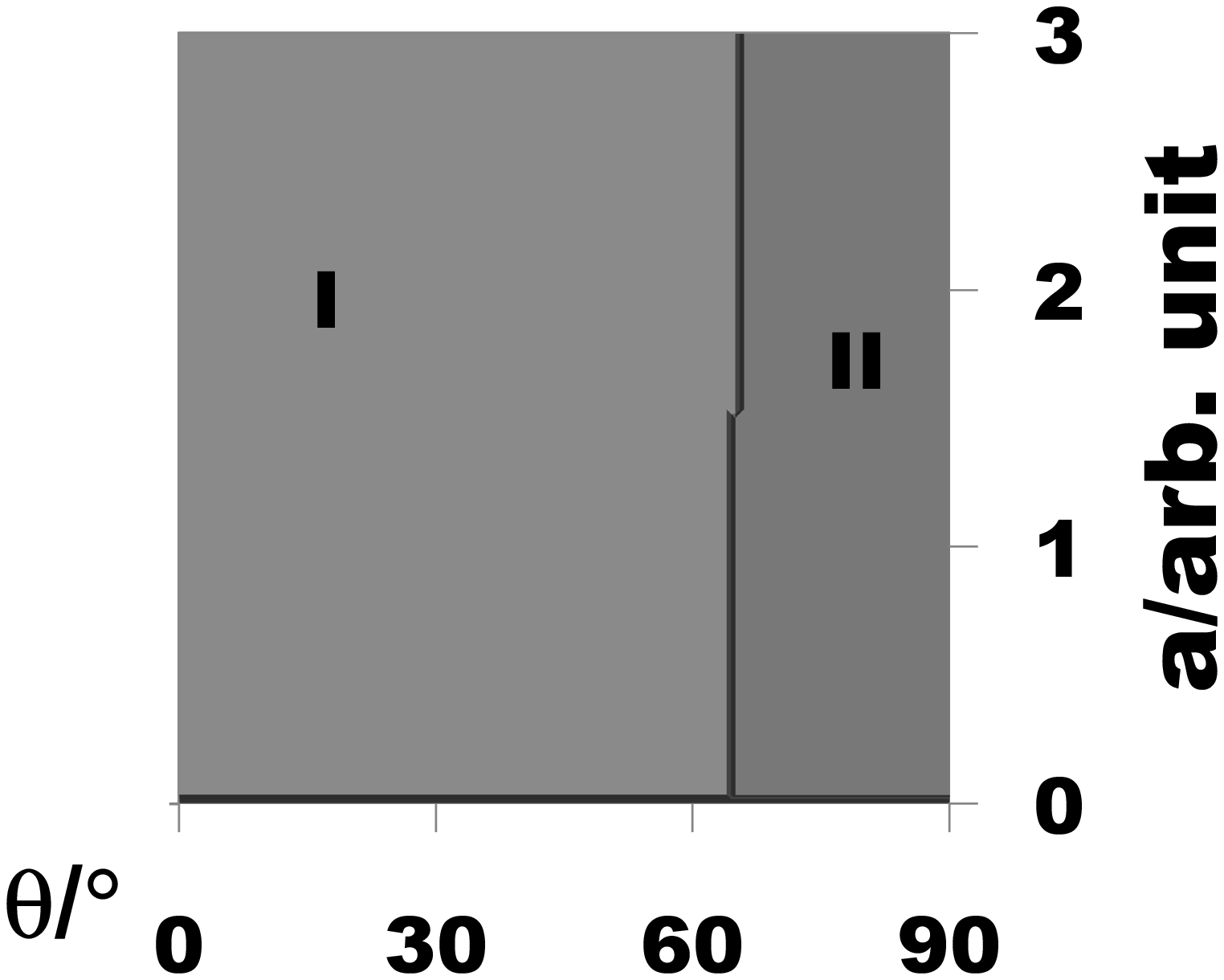}
\caption{\label{fig:sls_s2m6} Short length scale phase diagram for the anisotropic case
  $\sigma = 2$, $\mu = 6$ [top]; and vice-versa.}
\end{figure}

This implies that, as values of $\sigma$ and $\mu$ increases the
surface ripple patterns and 
their orientations, as defined by the regions, are observed with the
phase boundaries 
depending on $\theta$ and independent of the depth (Fig.\
\ref{fig:isos7m7}). In the rest of the 
paper we present our results for the anisotropic case, $\sigma \ne
\mu$. In Figs.\ 
\ref{fig:lls_sgtm} and \ref{fig:lls_s6m2}, we
present the 
large length scale sputtering behaviour, which consists of the regions
I, II, and III as 
defined above in section 2. The short length scale behaviour,
discussed below, 
is presented in Figs.\ \ref{fig:sls_s1m2} and \ref{fig:sls_s2m6}. 

A comparison of the phase diagrams of Figs.\ \ref{fig:isos7m7} to \ref{fig:lls_s6m2}
indicates that the phase boundaries of region two centered around
$\theta \approx 31^\circ$ in Fig.\ \ref{fig:isos7m7} shift to higher
values of $\theta$ as the relative size of the longitudinal stragle
with respect to the lateral straggle parameter $\mu$ increase, and to
lower $\theta<31^\circ$ as the relative size of $\sigma$ decrease, 
 such that the dominance of the patterns of regions I  
extend to higher or lower sputtering angles, respectively.  With
decreasing relative size of $\sigma$, we found a shift of region I
upwards due to constraints imposed by the increasing dominance of the
topography defined by region III. 
Thus, surface patterns of region III now dominate almost the entire
morphology for 
experimentally significant incidence angles, and penetration
depths. And, topography defined 
by region I, or region II, is almost irrelevant for common
sputtering 
angles. 

According to Cuerno and Barabasi,\cite{Cuerno1995} the three regions,
found and presented 
thus far, are applicable to the large length scale properties of the
surface morphology, and 
indicative of the three scaling regions characteristic of sputtered
surfaces at such 
length scales. As pertains to ripple morphologies alone, therefore,
there are only 
two possible regions either one having an orientation that is
perpendicular to the other, as 
explained in section \ref{sec:theory}. Such ripple topographic regions
account for 
the short length scale behaviour. In the rest of the paper, we
restrict ourselves to 
this characterisation of the surface morphology in terms of ripple
topographies and present 
results for some anisotropic cases. The regions in the phase diagrams
presented below are defined as follows, with the usual orientation characteristics:
I. $\nu_x < \nu_y \le 0$; II. $\nu_y < \nu_x$.

In Figure \ref{fig:sls_s1m2} top, we observe that patterns on the surface
are more of the 
orientation defined by region I than that defined by II. For angles
less than $\theta_{C_1}\approx 54^\circ$, the ripple orientation is
definitely along 
the x-direction, and for incidence angles greater than
$\theta_{C_2}\approx 68^\circ$, the ripple orientation is along the
y-direction for this case of anisotropic collision cascade
geometry. For angles within 
the range $\theta_{C_1} \le \theta \le \theta_{C_2}$, the ripple
orientation 
depends on the angle of incidence and penetration depth of the
impinging ion 
within the substrate as in the figure. 
This interpretation of the result of Fig. 8 applies to Figs.\
\ref{fig:sls_s1m2} (bottom) and \ref{fig:sls_s2m6}, in which cases we also 
have anisotropic geometries of the ion straggle. We will just note
here that the 
difference is due to the anisotropy, and that changes in the
boundaries shown in 
the figures are due to the relative magnitudes of the cascade
parameters as 
explained above. Note the saturation behavior displayed in the 
phase diagrams of Fig.\ \ref{fig:sls_s2m6}.

\section{\label{sec:conclusions} Conclusions}
We have performed phase diagram calculations for practical anisotropic
colision 
cascade parameters yet unreported in order to gain clearer insight
into the 
viewpoint of the continuum theory as regards the origin and
theoretical explanation 
of recent nanodot topographies observed for off-normal incidence
sputtering. In any 
case the anisotropic inclosing geometry of the ion straggle within the
substrate is 
the most likely in reality but unreported as presented here largely
because the 
isotropic cases that have been reported so far have been mainly for
simplicity and 
ease of presentation of the theory. In this first instance, we have
considered the three 
scaling regions defined by Cuerno and Barabasi in their introduction
of the 
widely acceptable form of the continuum theory. 

We found shifts in these phase boundaries as the relative magnitude of
the collision cascade 
parameters vary, with saturation for higher values of these
parameters. The saturation 
behaviour has been found for values of
the collision 
cascade parameters higher than the penetration depth. Specifically, we 
observed that when the relative size of the longitudinal straggle
$\sigma$ is greater than unity, the patterns and large length-scale
scaling behaviour of 
region I dominates while at the same time the phase boundaries
straighten from 
their curved outline until saturation, with increasing $\sigma$. On
the other hand, 
when the relative size of $\sigma$ is less than unity, patterns (and
orientation) and 
scaling of region III dominates while at the same time the influence
of region I reduces 
with increasing $\mu$. The saturation behaviour of the phase
boundaries is also found 
in this case. 

This exchange of patterns and orientations strongly suggests a
superposition 
of several patterns and orientation rapidly or gradually (as the case
may be) 
occurring on the topmost surface layer downwards. This may be the
factor 
responsible for the recently reported nanodot topographies predicted
for 
oblique incidence ion sputtering. Hence, while the results are not yet
conclusive 
about nanodot formation, they at least indicate that the nanodot
topographies may 
not constitute a breakdown of the continuum theory.

Through our comparative study of the weighting factors in the
stochastic partial differential equation that describes the surface
evolution, we 
have been able to provide these topographic phase diagrams predictive
of the 
large length scale and the short length scale behaviour to be expected
in 
experimental investigations that will almost always obtain anisotropic
collision 
cascades in their surface sputtering. Experimental investigations of
the scaling 
and topographic regimes reported here will constitute an important
step in an 
understanding of the scaling exponents characterizing the large length scale regions.

\begin{acknowledgments}
EOY thanks Alexander Hartmann, Reiner
Kree, and Rodolfo Cuerno for discussions while at G\"ottingen.  
\end{acknowledgments}

\providecommand{\noopsort}[1]{}\providecommand{\singleletter}[1]{#1}%

\end{document}